# Scenarios about the long-time damage of silicon as material and detectors operating beyond LHC collider conditions[1]

**I. Lazanu[a)], S. Lazanu[b)]**

[a)] University of Bucharest, Faculty of Physics, POBox MG-11, Bucharest-Magurele, Romania
e-mail: i_lazanu@yahoo.co.uk

[b)] National Institute for Material Science, POBox MG-7, Bucharest-Magurele, Romania,
e-mail: lazanu@alpha1.infim.ro

**Abstract**

For the new hadron collider LHC and some of its updates in luminosity and energy, as SLHC and VLHC, the silicon detectors could represent an important option, especially for the tracking system and calorimetry. The main goal of this paper is to analyse the expected long-time degradation in the bulk of the silicon as material and for silicon detectors, in continuous radiation field, in these hostile conditions. The behaviour of silicon in relation to various scenarios for upgrade in energy and luminosity is discussed in the frame a phenomenological model developed previously by the authors. Different silicon material parameters resulting from different technologies are considered to evaluate what materials are harder to radiation and consequently could minimise the degradation of device parameters in conditions of continuous long time operation.

---

[1] Work in the frame of CERN RD-50 Collaboration

# 1. Introduction

Particle physics makes its greatest advances with experiments at the highest energies. The way to advance to a higher energy regime is through hadron colliders, until now the Tevatron, in the near future the Large Hadron Collider (LHC) and beyond that its upgrades, the Super-LHC (SLHC) and the hypothetical Very Large Hadron Collider (VLHC).

After LHC, both a luminosity upgrade, with a factor of 10, up to $L \cdot 10^{35}$ cm$^{-2}$s$^{-1}$, and an energy upgrade, with a factor of 2, up to 28 TeV are being discussed (SLHC). An example of a design study for a Very Large Hadron Collider capable of reaching energies of order 240 TeV has recently been presented [1], [2]. The study supposes a two stages pp collider: in the first phase, a total centre of mass energy of 40 TeV and a luminosity of $L \cdot 10^{34}$ cm$^{-2}$s$^{-1}$ are expected, followed by the second stage when the energy would be at values up to 240 TeV and the luminosity of the order $L \cdot 10^{35}$ cm$^{-2}$s$^{-1}$. This proposed machine will be housed in a tunnel with a total circumference of 233 Km and is different from the older Eloisatron project.

The physics potential of these future hadron colliders has been discussed in the last years. Details about the possible scenarios which would dictate exploration at high energies up to 240 TeV could be found in references [3], [4], [5].

The LHC will start running in 2007 and at the present time the LHC detectors are well into their construction phase; silicon detectors represent an important option especially for the tracking system and calorimetry.

The investigation of the detector requirements for the next hadron colliders is necessary. At the present moment, systematic and detailed studies do not exist for future colliders, except those referring to LHC. The main goal of this paper is to analyse the expected long term degradation of the silicon material and of silicon detectors in the extreme hostile conditions expected to the future accelerators. The behaviour of silicon and of silicon detectors in relation to various scenarios for upgrade in energy and luminosity at SLHC and VLHC is discussed. In the same time, different silicon material parameters resulting from different technologies are considered to evaluate what materials are harder to radiation and consequently could minimise the degradation of device parameters in conditions of long time operation.

In the present paper, complex defect production and annealing are investigated in the frame of a phenomenological model, previously developed by the authors. The macroscopic modifications induced in detectors (leakage current and effective carrier concentration in the space charge region) are estimated on this basis.

# 2. Scenarios for radiation field at future colliders: LHC, SLHC, VLHC

The Large Hadron Collider at CERN, proposed in 1984, is expected to operate from 2007. The main research goals of physics, to be realised in two major experiments, ATLAS and CMS, have intensively been discussed; see for example Refs. [6], [7], [8], [9], [10]. Despite the technological difficulties, significant upgrades of the accelerator in energy and luminosity are considered as Super-LHC and Very-LHC respectively. The upgrade path will be defined by the results from the initial years of LHC operation.

At the present time the radiation fields in approximate experimental configurations are estimated only for LHC. At LHC energies, for bunch spacing of the order 25 ns and a luminosity of the order of $L \cdot 10^{34}$ cm$^{-2}$s$^{-1}$, the minimum bias events are assumed to have $n_{ch} \cong 6 \div 7$ particles per unit of pseudorapidity, and the average value of transverse momentum is $<p_T> = 0.45 \div 55\ GeV/c$. Inside the tracking cavity the hadrons represent around 54% of all the produced particles, and charged pions are the most abundant, around 64% of all hadrons [11]. In the concrete case of the CMS future experiment the simulated charged hadron spectra at different positions inside the tracking cavity have been published in Ref. [8]. The shape of the spectra as well as the maximum in particle distributions is dependent on the position inside the tracking cavity. In the spectra, the main contribution is due to pions, followed by protons, other hadrons being irrelevant in the distributions. The two positions considered are: a) r = 20 cm, z = 0÷60 cm, which corresponds to the maximum in flux, and b) r = 100 cm, z = 140÷280 cm; associated with the minimum hadron fluxes; from Ref. [8].

For LHC upgrades, in the absence of detailed studies, only suppositions are possible. Following the idea exposed by F. Gianotti in Ref. [5], we supposed the following conditions for the SLHC and VLHC respectively:

For SLHC environments, the following scenarios are considered:
      a) the pion and proton spectra remain the same as in LHC conditions but with one order of magnitude increase in intensity, corresponding to the order of magnitude increase of luminosity (no upgrade in energy considered),
      b) the luminosity is increased as in a), but the beam energy is increased with a factor of two, the energetic distribution of pions and protons is the same as for LHC conditions, but the average energy of the spectra is shifted to higher energy with 50 MeV;

For the upgrade to VLHC, in the estimation of the generation rate we supposed:
      c) the same geometrical configurations as for LHC, the same distributions of particles in the corresponding positions in the tracking cavity, but with the maximum in the spectra shifted to higher energies with about 150 MeV at the same luminosity,
      d) one order of magnitude increase in the luminosity, in respect to c), respectively.

## 3. Model for long time kinetics of defects in irradiated silicon

In this work, the effects of irradiation conditions and various initial impurities in the starting material are discussed in the frame of the phenomenological model developed by the authors and discussed extensively in previous papers [12], [13], [14], [15]. In short, the model supposes three steps.

In the first step, the incident particle interacts with the semiconductor material. The peculiarities of the interaction mechanisms are explicitly considered for each kinetic energy.

In the second step, the recoil nuclei resulting from these interactions lose their energy in the semiconductor lattice. Their energy partition between displacements and ionisation is considered in agreement with Lindhard's theory [16] and authors' contributions [17].

A point defect in a crystal is an entity that causes an interruption in the lattice periodicity. The terminology and definitions in agreement with M. Lannoo and J. Bourgoin [18] are used in relation to defects. We denote the displacement defects, vacancies and interstitials, as primary point defects, prior to any further rearrangement. After this step the concentration of primary defects per unit particle fluence (CPD) is calculated. The generation term ($G$) is the sum of two components: $G_R$ accounting for the generation by irradiation, with defects supposed randomly distributed, and $G_T$, for thermal generation. In silicon, vacancies and interstitials are essentially unstable and interact via migration, recombination and annihilation or produce complexes defects by self interaction or interacting with impurities.

Some comments are necessary. CPD is calculated considering an average energy threshold for displacements and consequently eliminating the influence of the anisotropy of the lattice on defect production. Because the incident particle has, in the cases of interest for the present study, high energy and the silicon detector is a thin target, this particle loses only a small fraction of its energy in the detector by nuclear interactions. In Lindhard's theory, after this single interaction, the identity of the primary incident particle is lost, and its identity remains only correlated with the energy transferred to the recoil, and subsequently with the number of primary defects. Thus, two different particles, with different energies, could produce the same rate of generation of primary defects after a nuclear interaction if the following condition is fulfilled:

$$[CPD(E_1) \times Flux(E_1)]_{particle\ 1} = [CPD(E_2) \times Flux(E_2)]_{particle\ 2} \quad (1)$$

The production of complexes of defects or clusters could follow, depending on the values of CPD, number and species of impurities and the conditions of stability for these new defects.

The concentration of primary defects represents the starting point for the third step of the model, the consideration of annealing processes, treated in the frame of the chemical rate theory. A review of possible mechanisms of complex defects formations in silicon, consequence of irradiation processes could be found, e.g. in Reference [12].

Without free parameters, if the initial impurity concentrations and the chemical rates of processes are known, the model is able to predict the absolute values of the concentrations of defects and their time evolution toward stable defects, starting from the primary incident particle characterised by type and kinetic energy.

In previous papers cited above, the formation and evolution of defects in silicon, around room temperature, has been described as follows:

$$V + I \xrightarrow{K_1}_{G} 0 \quad (2a) \qquad V + V \underset{K_5}{\overset{K_4}{\rightleftarrows}} V_2 \quad (3a)$$

$$I \xrightarrow{K_2} sinks \quad (2b) \qquad I + V_2 \xrightarrow{K_6} V \quad (3b)$$

$$V \xrightarrow{K_3} sinks \quad (2c)$$

$$V + P \underset{K_7}{\overset{K_4}{\rightleftarrows}} VP \quad (4a) \qquad I + VP \xrightarrow{K_8} P \quad (4b)$$

$$V + O \underset{K_9}{\overset{K_4}{\rightleftarrows}} VO \quad (5a) \qquad I + VO \xrightarrow{K_{10}} O_i \quad (5b)$$

$$I + C_s \xrightarrow{K_1} C_i \quad (6a) \qquad V + C_i \xrightarrow{K_{11}} C_s \quad (6b)$$

$$V + VO \underset{K_{12}}{\overset{K_4}{\rightleftarrows}} V_2O \quad (7a) \qquad I + V_2O \xrightarrow{K_{13}} VO \quad (7b)$$

$$C_i + C_s \underset{K_{15}}{\overset{K_{14}}{\rightleftarrows}} C_iC_s \quad (8)$$

$$C_i + O_i \underset{K_{16}}{\overset{K_{14}}{\rightleftarrows}} C_iO_i \quad (9)$$

where the reaction constants $K_i$ (i = 1, 4 ÷ 16) have the general form: $K_i = C \cdot v \cdot \exp(-E_i/k_BT)$, with $v$ the vibration frequency of the lattice, $E_i$ the associated activation energy and $C$ a numerical constant that accounts for the symmetry of the defect in the lattice. The reaction constants related to the migration of interstitials and vacancies to sinks could be expressed as: $K_j = \alpha_j \cdot v \cdot \lambda^2 \cdot \exp(-E_j/k_BT)$, with j = 2 (interstitials) and 3 (vacancies), $\alpha_j$ - the sink concentration and $\lambda$ - the jump distance.

The activation energies associated with processes 2, 3a, 4a, 5a, 6a, 7a, 8 and 9 have been taken in agreement with the values reported in the literature: $E_1 = E_2 = 0.4$ eV, $E_3 = E_4 = 0.8$ eV, $E_5 = 1.3$ eV, $E_7 = 1.1$ eV, $E_9 = 1.4$ eV, $E_{12} = 1.6$ eV, $E_{14} = 0.8$ eV, $E_{15} =$, $E_{16} = 2$ eV. For process 5b, the activation energy is $E_{10} = 1.7$ eV, in agreement with the value reported experimentally [19]. Due to the fact that most of the activation energies associated to reactions of decomposition of vacancy-impurity complexes, initiated by interstitials, are not available from the experimental literature, the other 3 activation energies are introduced as parameters; the following numerical values have been assigned to them: $E_6 = E_{10} = E_{11} = 1.7$ eV; $E_8 = 1.5$ eV, and $E_{13} = 2$ eV.

Complexes of other impurities with primary defects could also be added to this reaction scheme. The existence of clusters is not considered in the present kinetics model. Until now, there exist more ambiguities in relation to cluster characteristics, their experimental identification and their macroscopic effects. Most probably the cluster should be regarded as one big defect that could introduce a high number of electronic states into the bandgap. The number of electrons or holes it could trap is limited to quite a few because of carrier repulsion.

A special discussion must be made for process (2c). It has been found [12] that the migration of vacancies to sinks could be neglected, due to the fact that five orders of magnitude separate the diffusion constant of interstitials and vacancies respectively in silicon at room temperature, this being a direct consequence of the fact that the energy of migration of the vacancy is significantly higher than that of interstitials.

## 4. Parameters of non-irradiated silicon for detectors for HEP environment

Traditionally, the silicon sensors have been fabricated using the float zone (FZ) crystal growth technique. For detector applications, the float zone technique ensures high-pure and defect-free silicon. Due to the high resistivity of the material, the detector can be fully depleted at relatively low voltage. Naturally the silicon obtained by FZ growth is characterised by low oxygen concentration. At BNL [20], a technique to incorporate oxygen in the bulk of the material to improve the radiation hardness of silicon has been developed. For this material we refer as DOFZ (or alternatively HTLT).

Usually, silicon obtained by the Czochralski growth technology (Cz-Si) is characterised by low resistivity. In the last period, silicon with sufficient high resistivity (MCz Magnetic Czochralski) permitting to obtain detectors is being fabricated.

In the analysis that follows, only the presence of phosphorus, oxygen and carbon impurities is considered in the silicon bulk, and the impurities are supposed to be uniformly distributed. For high resistivity silicon, phosphorus concentrations between $4 \times 10^{11} \div 1 \times 10^{12}$ $atoms/cm^3$ have been used, for medium resistivity $3 \times 10^{12} \div 5 \times 10^{12}$ $atoms/cm^3$ and for low resistivity $1 \times 10^{13}$ $atoms/cm^3$ respectively. The carbon concentrations are considered to be in the range between $1 \times 10^{15}$ $atoms/cm^3$ and $1 \times 10^{17}$ $atoms/cm^3$. The content of oxygen is correlated with the growth technology. In the FZ technique the oxygen concentration is about $1 \times 10^{15}$ $atoms/cm^3$ and this concentration attains values around $(1 \div 4) \times 10^{17}$ $atoms/cm^3$ in DOFZ Si. For Cz-Si, we used concentrations of oxygen in the range $2 \times 10^{18} \div 4 \times 10^{18}$ $atoms/cm^3$ and between $8 \times 10^{17} \div 1 \times 10^{18}$ $atoms/cm^3$ for MCz technology respectively.

## 5. Predictions for material structural time modifications and detector change parameters operating in LHC, SLHC and VLHC conditions

In the LHC conditions (considering in the concrete discussions the particular case of CMS experiment) the hadrons are the predominant particles in the tracker, especially low energy charged pions and protons. The maximum in the rates of primary defects generated by pions comes from the region around 200 MeV while for protons the major contribution comes from the lowest energy region, see for example Ref..[21]. In the concrete calculations, the CPD distribution induced by pions is cut at 20 MeV. This cut represents only a contribution below 0.5% in the integrated defect spectra. The differential energetic generation rates of defects for SLHC and VLHC conditions are calculated in accord with the hypotheses discussed in paragraph 3.

In Figure 1 the differential energetic generation rates of primary defects, from left to right, for the LHC, SLHC and VLHC environment, for protons (triangles) and pions (asterisk and rhombus), for two extreme possible positions in the tracking cavity are represented. The SLHC and VLHC differential generation spectra have been obtained by increasing the average energy with 50 MeV and 150 MeV respectively. The increase in luminosity is not included.

In the two extreme radiation fields existent in the tracker cavity and considered in this work, the rates of generation of defects (induced by pions and protons) are: 6.8 x 10$^8$ VI/cm$^3$/s in the LHC conditions and 6.8 x 10$^9$ VI/cm$^3$/s, (SLHC, hypothesis a)), 7.2 x 10$^9$ VI/cm$^3$/s (SLHC, hypothesis b)), 6.9 x 10$^8$ VI/cm$^3$/s for VLHC, hypothesis c) and 6.8 x 10$^9$ VI/cm$^3$/s, VLHC, hypothesis d) respectively. For these generation rates, the contributions coming from proton spectra represent 8.2%, 8.2%, 6%, .8% and 8% respectively.

In the considered scenarios, the maximum of the generation rate of primary defects is obtained in the hypothesis b) for the SLHC upgrade. VLHC conditions are less dangerous for silicon detectors that the SLHC environment.

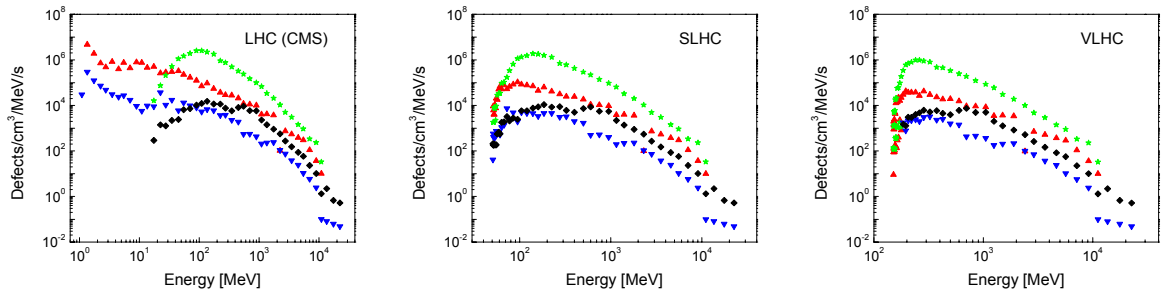

**Figure 1.**

Differential energetic generation rates of primary defects, from left to right, in LHC, SLHC and VLHC environment, for protons (triangles) and pions (asterisk and cross), for two extreme possible positions in the tracking cavity, r=20 cm, z=60÷120 cm and r=100 cm, z=140÷280 cm respectively, in accord with hypothesis b) and c) – see text.

† **Estimated concentration of stable defects and their time evolution in the LHC, SLHC and VLHC conditions**

The time dependencies of the concentrations of $V_2$, $V_2O$, VP, VO and $C_iO_i$ are relevant for the long time material degradation in the conditions corresponding to LHC and SLHC respectively. In this analysis four possible materials have been considered, with the characteristics listed in Table I.

**Table I.** Silicon characteristics used in the calculations

| Material | [P] [atoms/cm$^3$] | [O] [atoms/cm$^3$] | [C] [atoms/cm$^3$] |
|---|---|---|---|
| FZ | $4 \cdot 10^{11}$ | $10^{15}$ | $10^{16}$ |
| DOFZ | $4 \cdot 10^{11}$ | $4 \cdot 10^{17}$ | $10^{16}$ |
| MCz | $3 \cdot 10^{12}$ | $8 \cdot 10^{17}$ | $10^{16}$ |
| Cz | $10^{13}$ | $4 \cdot 10^{18}$ | $10^{16}$ |

The resulting dependencies are represented in Figure 2 as a function of the irradiation time.

From the examination of these curves, it could be observed that the concentration of divacancies is higher in low resistivity FZ silicon, which has small concentrations of impurities which could trap vacancies, both for LHC and SLHC conditions. One order of magnitude increase of the irradiation rate, corresponding to the upgrade, conduces to an important increase of [$V_2$] after some years of operation both for DOFZ and MCz materials. In Cz-Si a smooth time increase of [$V_2$] is predicted both for LHC and for SLHC, but one order of magnitude increase in the generation rate is reflected in nearly two orders of magnitude higher [$V_2$].

The concentration of the most dangerous defect, $V_2O$, increases with the irradiation rate, but attains a plateau both for FZ and DOFZ-Si at SLHC, being limited by the oxygen content. One could note that [$V_2O$] is the highest in DOFZ silicon for LHC conditions, although the oxygen content of DOFZ is lower than oxygen concentration in Cz and MCz-Si, which have both higher P. Phosphorus addition could be a clue in minimising the [$V_2O$], but this must be discussed in balance with the full depletion voltage optimisation. [$V_2O$] is the lowest in FZ silicon, being limited by the oxygen content of the material, and attains the corresponding plateau after some months of operation at LHC and faster at SLHC. The plateau defined by the oxygen concentration is attained also for DOFZ materials in SLHC conditions, after around 5 years of operation.

In what regards the VP defect, its concentration is not directly determined by [P], Cz-Si having in both irradiation conditions relatively low concentrations of VP. It is important to note also that in the situations examined here, P does not transform integrally by capturing vacancies into VP.

The production of VO is practically identical in Cz-Si, MCz-Si and DOFZ in the conditions of irradiation rates expected to LHC, and completely different for FZ-Si: a pronounced decrease in time followed by saturation.

This step decrease in time of [VO] is observed in SLHC conditions for all materials, except MCz-Si, where a plateau similar to the one found in LHC conditions are predicted.

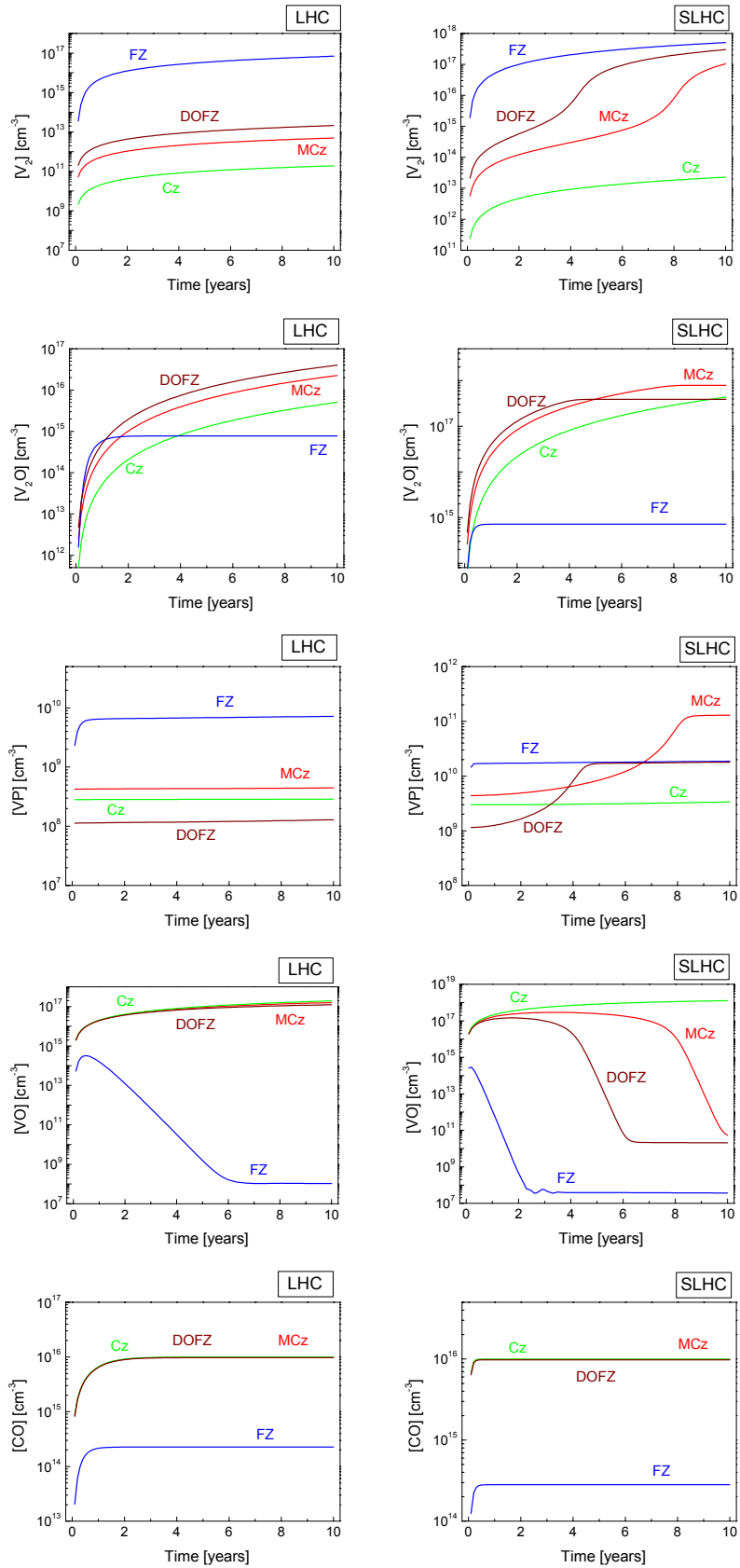

**Figure 2.**

Time evolution of the concentration of complex defects in the LHC and SLHC conditions

[C$_i$O$_i$] concentrations are the lowest in FZ Si both for LHC and for SLHC, and have similar values for all the other materials.

In the range of rates considered in the present paper, the content of carbon in silicon (between $10^{15}$ and $10^{17}$ C/cm$^3$) has the greatest influence on FZ high resistivity materials, influences DOFZ ones and has little or no influence on Cz-Si and MCz-Si. From the defects, the most sensitive to the carbon content are the interstitials, the interstitial carbon, C$_i$O$_i$ and C$_i$C$_s$. The concentration of [C$_i$O$_i$] and [C$_i$C$_s$] increases monotonically with the concentration of C. The other defects are indirectly influenced by the amount of carbon present in the sample, having variable behaviours as a function on the P and O concentrations, and on the irradiation rate. For example, [V$_2$O] in DOFZ has the highest values for carbon concentration around $10^{17}$ atoms/cm$^3$ in LHC conditions, and the lowest values for SLHC rates. The divacancy concentration increases monotonically with the carbon concentration, while [VP] has at low irradiation times a peak for high carbon materials.

### † Regularities predicted for defect concentrations in different materials and for different irradiation rates

In the present theoretical investigation, regularities between different defect concentrations have been observed, but due to the great variability of material compositions and irradiation rates only particular situations are discussed.

At low generation rates and high oxygen concentration in the starting material, the model predicts a ratio of [VO] to [VP] independent on the irradiation rate, and with linear time dependence. The range of rate for which this phenomenon exists increases with the oxygen concentration, e.g. in FZ-Si this is 1.4·$10^6$ VI/cm$^3$/s, in OFZ - 7 $10^6$ VI/cm$^3$/s, and in Cz - 7 $10^7$ VI/cm$^3$/s.

This observation is in good agreement with the experimental result of Su [22], who found for gamma irradiated Cz Si linear dependence of [VO] on [VP].

The ratio [V$_2$]/[VP] normalised to the irradiation rate does not depend on the irradiation rate and is linear in time at low irradiation rates and high oxygen concentration. This feature is maintained independently on the resistivity (phosphorus concentration) in the starting material. Again, the increase of the oxygen concentration extends the domain of rate for which the phenomenon is valid. For $10^{15}$ O atoms/cm$^3$, the maximum rate is 2 $10^6$ VI/cm$^3$/s, for $10^{16}$ O/cm$^3$ it is 2 $10^7$ VI/cm$^3$/s, while for $10^{17}$ O/cm$^3$ - 2 $10^8$ VI/cm$^3$/s.

The experimental data of Brotherton [23] for electron irradiated Cz-Si for [V$_2$] as a function on [VO] for different irradiation rates are supported by the model predictions.

### † The problem of dimers and other mechanisms

The study of oxygen dimers has been initiated in different laboratories in the hope to find new mechanisms by which the radiation hardness of silicon could be increased [24].

In the present paper, dimers have been considered for the estimation of their possible influence on defect production and time evolution in silicon in the LHC/SLHC conditions, and also to compare model prediction with experimental data existent in the literature.

In irradiated silicon, VO and V$_2$O are produced by the capture of moving vacancies by interstitial oxygen, and by the VO centre respectively. While the VO centre is charged only at cryogenic temperatures, the V$_2$O defect is charged up to 90% already at room temperature and therefore produces detrimental effects on the macroscopic detector parameters. A possible solution to diminish these uncalled effects is to produce dimers in silicon which can act as a sinks for migrating interstitials by forming IO$_{2i}$ and I$_2$O$_{2i}$ defects. J. L. Lindström and co-workers [25], discussed the importance of the interactions between the oxygen dimer ($O_{2i}$) and silicon self-interstitials and vacancies. In Cz-Si technology oxygen dimers are produced at least by two mechanisms: during the crystal growth, when their concentration is around $1 \times 10^{15}\ cm^{-3}$, and/or during high temperature irradiation. In this paper we discuss only the processes taking place at room temperature. The main reactions induced by dimers in these conditions are:

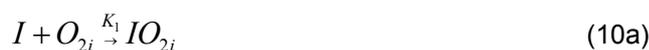

$$I + O_{2i} \xrightarrow{K_1} IO_{2i} \qquad (10a)$$

$$I + IO_{2i} \xrightarrow{K_1} I_2O_{2i} \qquad (10b)$$

Both reactions are initiated by self-interstitials, and their rates are related to the activation energy of interstitials.

The $IO_{2i}$ centre is electrically active with an acceptor level at $E_C - 0.11\ eV$. Both $IO_{2i}$ and $I_2O_{2i}$ are stable at room temperature and anneal out at about 400 K and 550 K, respectively.

If in the material there exist VO$_2$ centres (produced previously by irradiation at higher temperature, but which are stable at room temperature), they are also capturing interstitials, producing oxygen dimers by the reaction:

$$I + VO_2 \xrightarrow{K_1} O_{2i} \qquad (10c)$$

This reaction is also produced by the migration of interstitials, and has the same reaction constant as reactions (10a) and (10b).

Model predictions have been compared with the experimental data published by Lindstroem in Ref [25], where the effect of electron irradiation at room temperature on a Cz-Si sample containing both VO$_2$ centres and oxygen dimers, in the concentration range 4 ÷ 5 x 10$^{16}$ cm$^{-3}$, is investigated.

In Figure 3, the fluence dependencies of the concentrations of O$_{2i}$, VO$_2$, IO$_2$ and I$_2$O$_2$ predicted by the present model, supplemented with equations 10a, 10b and 10c are compared with experimental data after electron irradiation at room temperature of a dimer-rich Cz-Si sample.

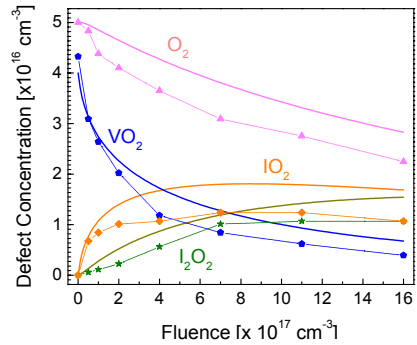

**Figure 3.**

The kinetics of complexes associated with dimers in Cz-Si at RT: model calculations – continuous line and normalised experimental data - points and dotted lines from Ref. [25].

Due to the incomplete information regarding irradiation and measurement conditions in the published paper, the comparison is only at the qualitative level.

A good agreement between model predictions and experimental data has been obtained.

Due to the fact that as-grown Cz-Si contains oxygen dimers in the order 10$^{15}$ cm$^{-3}$, mechanisms 10a and 10b have been introduced into the reaction scheme in the study of defect evolution of Cz and MCz-Si in LHC and SLHC conditions. Negligible contributions of these two mechanisms have been found in these conditions of irradiation and initial concentration of dimers.

### ü Effects at the detector level: leakage current and the effective concentration in the space charge region

Most of the radiation detectors are based on the properties of the p-n junction. Consequences of irradiation processes, some characteristics of the devices could be influenced by the formation of secondary defects which are unstable and which could be electrically active - with energy levels located in the band gap of silicon. In this case, they could capture free electrons or holes and change the initial concentrations of charged donors and acceptors in the space charge region of the detector. The principal effects are: change of depletion voltage, increase of leakage current, bulk material resistivity modification, change of electric field distribution in irradiated silicon p-n junction, charge in the collection efficiency and capacitance contribution to the noise.

In this paper, only some aspects related to the increasing of the leakage current and to the modification of the effective concentration of donors and acceptors will be discussed. The changes of the effective concentration of donors and acceptors ($N_{eff}$) in time during irradiation are responsible for the modification of the depletion voltage.

The defects considered in the model, relevant for the modifications of leakage current and for the effective concentration are those with deep energy levels in the band gap, in the neighbourhood of the intrinsic level: these are the deepest level of the divacancy, VP, the level associated with the $C_iO_i$ centre, and the level at $E_c - 0.52$ eV, assigned by us to $V_2O$. Other defect levels, put in evidence by experiment but not assigned to defect centres [26], could be important and could modify the present predictions, but are not considered.

The leakage current is evaluated in the frame of the simplified Shockley-Read-Hall model. One must specify that generally the calculation of the leakage current starting from concentrations of defects conduces to discrepancies, more important for hadron irradiation [27], [28] possible sources are discussed in a previous paper [21].

To diminish these discrepancies, in Figure 4 the ratio of leakage currents estimated for SLHC and LHC conditions is presented in the same materials for which defect concentrations have been calculated. Short time after the beginning of the irradiation, an important increase of this ratio is obtained. The predictions indicate that in FZ Si, an increase of one order of magnitude in the irradiation rate in respect to LHC produces a 5 times higher current in detectors, practically constant in time. DOFZ behaves worse, MCz even worse and Cz Si the worst.

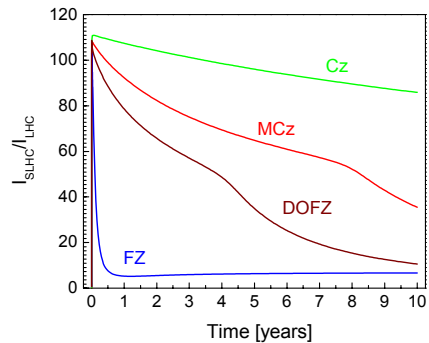

**Figure 4.**

Time dependence of the ratio between the leakage current in SLHC and LHC conditions.

In Figure 5 the changes of the effective carrier concentration in the space charge region of the detector are presented as a function of the irradiation time. These dependencies are similar with the dependencies of $N_{eff}$ on the irradiation fluence, for charged and neutral hadrons, presented in the literature (see, e. g. Ref. [24] and papers cited therein). However, a direct correspondence couldn't be done between these curves because the present calculations refer to continuous irradiation, with a complex spectrum of particles, while in the literature post irradiation conditions are reported, at similar fluences, but obtained from different irradiation rates, and usually with mono energetic particles.

In LHC conditions, a beneficial effect of the oxygenation of FZ Si on $N_{eff}$ is predicted by the model, oxygen addition producing a delayed inversion of the structure in the same conditions of initial phosphorous content. At longer times, the FZ material proves to be the best choice, the increase of $N_{eff}$ being limited to $4 \times 10^{13}$ cm$^{-3}$, while in DOFZ such a limitation has not been found. The same beneficial effect of oxygen on delayed inversion could be remarked in Cz-Si in respect to MCz.

In conditions of higher rates, at SLHC, all materials invert faster, in times of the order 0.1 years and in all materials, after 10 years of operation, higher values of $N_{eff}$ have been found in the frame of the model in comparison to LHC. This time, DOFZ inverts before FZ, but in FZ-Si $N_{eff}$ is the smallest from all materials investigated at longer times, after inversion.

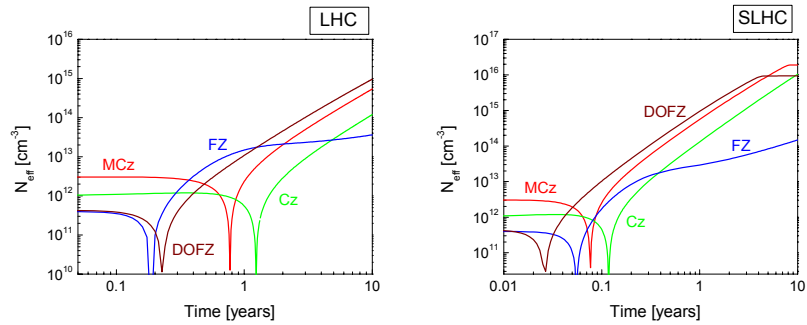

**Figure 5.**

Time dependence of the changes of the effective carrier concentration in the space charge region of the detector

The phenomena of trapping and de-trapping of charge carriers on defect levels are responsible for the decrease of the charge collection efficiency (CCE): if the charge carrier, produced by the ionising radiation, trapped on a deep level is re-emitted at a time much later than the maximum collection time, it has to be considered as lost. These phenomena, non-investigated in the present paper, could be an important limitation for the technology at bunch spacing of the order $12.5 \div 18$ ns.

## 6 Possible conclusions

In the hypothesis considered in the paper, considering upgrades of LHC in energy and in the luminosity, the SLHC conditions are more dangerous for silicon detectors than LHC or VLHC environments.
From the point of view of the parameters analysed, excepting the first months of operating, the FZ technology is found to be, in the frame of the model, more adequate for the proposed scope.
Oxygen dimers, in concentrations corresponding to as-grown Cz-Si, do not enhance the radiation hardness of silicon.